\documentclass[epj]{svjour}

\usepackage{graphics}
\usepackage{amsmath,bm}
\usepackage{amssymb}
\usepackage{color}

\usepackage{epstopdf}

\graphicspath{{./}} 

\begin{document}
\title{Phase and precession evolution in the Burgers equation}

\author{Michele Buzzicotti\inst{1}\and Brendan P. Murray\inst{2}\and Luca Biferale\inst{1}\and Miguel D. Bustamante\inst{2}
}                     


\institute{Department of Physics and INFN, University of 
Rome ``Tor Vergata'',
Via della Ricerca Scientifica 1, 00133, Rome, Italy. \and Complex and Adaptive Systems Laboratory, School of Mathematics and Statistics,
University College Dublin,
Belfield, Dublin 4, Ireland.}

\date{18 April 2016}
%
\abstract{ We present a phenomenological study of the phase dynamics of the one-dimensional stochastically forced Burgers equation, and of the same equation under a Fourier mode reduction on a fractal set. We study the connection between coherent structures in real space and the evolution of triads in Fourier space. Concerning the one-dimensional case, we find that triad phases show alignments and synchronisations that favour energy fluxes towards small scales --a direct cascade. In addition, strongly dissipative real-space structures are associated with entangled correlations amongst the phase precession frequencies and the amplitude evolution of Fourier triads. As a result, triad precession frequencies show a non-Gaussian distribution with multiple peaks and fat tails, and there is a significant correlation between triad precession frequencies and amplitude growth. Links with dynamical systems approach are briefly discussed, such as the role of unstable critical points in state space. On the other hand, by reducing the fractal dimension $D$ of the underlying Fourier set, we observe: i) a tendency toward a more Gaussian statistics, ii) a loss of alignment of triad phases leading to a depletion of the energy flux, and iii) the simultaneous reduction of the correlation between the growth of Fourier mode amplitudes and the precession frequencies of triad phases.  
} 
\maketitle

\let\thefootnote\relax\footnotetext{Postprint version of the article published on Eur. Phys. J. E (2016) 39: 34 doi 10.1140/epje/i2016-16034-5}

\section{Introduction}
\label{intro}
The physics of extended nonlinear dynamical systems is often characterised by fluctuations at different frequencies and/or wavelengths. In many cases it is key to disentangle the evolution of amplitudes and phases of such fluctuations \cite{Ott94,Cencini2010,Wein1965}.  In this paper we  consider the context of nonlinear partial differential equations (PDEs) describing the  interaction of oscillating Fourier modes (waves) with quadratic nonlinearities in general, and the application to the one dimensional Burgers equation: 
\begin{equation}
\frac{\partial u}{\partial t} + \frac{1}{2}\frac{\partial u^2}{\partial x} = \nu\frac{\partial^2 u}{\partial x^2} + f, 
\label{eq:burgers}
\end{equation}
where $u(x,t)$ is the space periodic velocity, $\nu$ is a positive parameter (viscosity) and $f(x,t)$ is the external forcing. The Burgers equation, see \cite{becreview} for a review, is a model for various nonlinear dissipative systems. It describes a variety of nonlinear wave phenomena such as acoustic waves and plasma physics \cite{Whitham74}. 
When driven by a random forcing it has applications in condensed matter problems like interface deposition and growth (see for instance \cite{bs95}). In absence of a driving force, the explicit solution provided by the Hopf-Cole method \cite{h50,c51} leads to non-trivial problems in the limit of vanishing viscosity, when random initial conditions are assumed. This is particularly important 
for the evolution of large-scale structures in the Universe \cite{Zeld70,Gurb_Saic_Shan89,She_Aur_Fisch92,Vergas94,Aurell97}. 
Burgers equation was originally conceived as a toy model for turbulence and it is frequently used as a testing ground for numerical schemes and as a training ground for developing mathematical tools to study Navier-Stokes turbulence and other hydrodynamical or Lagrangian problems \cite{Benzi2010,Biferale2005,Biferale2013,Biferale2004,w98,Bustamante2011}. \\
Burgers equation represents one of the simplest nonlinear partial differential equations known to display a non-trivial scaling of the velocity field correlation functions. Multiscaling is connected to the tendency to create shocks and consequently to increase negative velocity differences $\Delta_r u < 0$ and decrease positive ones $\Delta_r u > 0$, where $\Delta_r u = u(x+r,t) - u(x,t)$. Thus a strongly non-Gaussian probability distribution function (PDF) of $\Delta_r u$ is observed. The presence of strongly localised velocity jumps (shocks) in the real space is the fingerprint of Burgers's dynamics. We remark that shocks are the only structures in the flow able to dissipate energy in the limit of small viscosity. In other words, the energy flux across scales is absorbed only by a few strongly localised events in real space. The guiding motivation of this paper is to look for relations between the real-space multiscaling properties and the Fourier dynamics, a key problem also in Navier-Stokes turbulence \cite{frisch,pope}.


We will work almost exclusively with Fourier mode variables $\hat {u}_{k}(t)$ rather than the real periodic field $ u(x,t)$, where $u(x,t) =\sum_{k \in \mathbb{Z}} e^{i k x} \hat {u}_{k}(t)$ with $x$ the position in real space and $k$ the wavenumbers. Reality of $u(x,t)$ implies $\hat {u}_{-k}(t) = \hat {u}_{k}^*(t)$, where ${}^*$ denotes complex conjugation. When Galerkin truncations are considered, the range for the sum above is reduced: instead of summing over $k \in \mathbb{Z}$ we sum over a given subset: $k \in \mathcal{C}$, where $\mathcal{C} \subset  \mathbb{Z}.$ Each mode is indexed by an integer wavenumber $k$. The dynamical content of mode $k$ is given by its complex valued Fourier amplitude, $\hat u_k(t)$. Using this representation, the governing PDE can be decomposed into a set of ordinary differential equations (ODEs) which describe the individual evolution of each of the complex amplitudes. To further separate the variables of interest an amplitude/phase representation will be used: $\hat u_k(t) = a_k(t) e^{i \phi_k(t)}$, where  $a_k(t) = |\hat u_k(t)|$ (amplitude) and $\phi_{k}(t) = \arg [\hat{u}_{k}(t)]$ (phase).

In Burgers equations the nonlinearity is quadratic so interactions appear in \emph{triads} (groups of 3 modes). The {\it key} dynamical degrees of freedom consist of the modes' real amplitudes $a_k(t)$ along with the \emph{triad phases}, \emph{i.e.} the combinations:  
$${\varphi}_{k_1,\,k_2}^{k_3}(t) = {\phi}_{k_1}(t) + {\phi}_{k_2}(t) - {\phi}_{k_3}(t),$$
where the wavenumbers $k_1$,$k_2$,$k_3$ satisfy a `closed-triad' condition: $k_1 + k_2 = k_3.$ By simple counting, 
one obtains that many triad phases are not linearly independent; a maximal set of linearly independent triad phases is easily found once the set of wavenumbers $\mathcal{C}$ is known. From here on, we will call ``state space'' the collection of modes' real amplitudes along with a maximal set of linearly independent triad phases.

The motivation for studying the evolution of these key dynamical degrees of freedom is that they provide quantitative information about the energy exchanges taking place in the system. 
For example, it was shown in \cite{bustamante14} for the 2D barotropic vorticity equation with periodic boundary conditions that the triad phases ${\varphi}_{k_1,\,k_2}^{k_3}(t)$ not only oscillate in time but also \emph{precess} (or drift). This precession gives rise to a new frequency, the so-called precession frequency, which is approximately equal to the value of the zero-mode in time of $\dot{\varphi}_{k_1,\,k_2}^{k_3}(t)$. When the precession frequency of a given triad is equal to zero or coincides with a typical frequency of the system's oscillation, then there is a \emph{precession resonance}, characterised by strong energy transfers. To see how this works, the contribution from a given triad to the energy flux towards small scales (to be defined below in Section \ref{sec:Fractal_burg}) is proportional to the time/ensemble average $\langle a_{k_1}a_{k_2}a_{k_3} \sin ({\varphi}_{k_1,\,k_2}^{k_3})\rangle\,.$ Two resonant scenarios are possible regarding the phase behaviour. i) The triad phase simply oscillates near the value $\pi/2 \mod 2\pi$. We get a non-zero contribution to the energy transfer from this average because $a_{k_1}a_{k_2}a_{k_3} >0$ by definition. This corresponds to precession resonance with zero precession frequency: $\langle \dot{\varphi}_{k_1,\,k_2}^{k_3}\rangle \approx 0$. ii) The phase drifts with non-zero precession frequency, for example ${\varphi}_{k_1,\,k_2}^{k_3}(t) \approx \Omega \, t$. Thus, the contribution to the flux has the form \\$\langle a_{k_1}a_{k_2}a_{k_3} \sin (\Omega \,t)\rangle$, which will not vanish if $\Omega$ coincides with  a typical frequency of the time signal of $a_{k_1}a_{k_2}a_{k_3}$. This corresponds to precession resonance with non-zero precession frequency: $\langle \dot{\varphi}_{k_1,\,k_2}^{k_3}\rangle \approx \Omega \neq 0$. Both resonant scenarios can involve multiple triads in synchrony, resulting in strong energy transfers across multiple scales of the system. This phenomenon leads naturally to the study of invariant manifolds within the state space. In \cite{bustamante14}, precession resonances were shown to correspond dynamically to trajectories embedded in the unstable manifolds of periodic orbits. These unstable manifolds have components that extend towards large values of amplitudes, $a_k(t)$, contributing directly to intense energy-transfer events.

The aim of this study is twofold. First, we want to study the relevance of triad phases and their precession frequencies relating to the energy transfers amongst modes, focusing on: \noindent i) Probability distribution of triad phases, joint probability distributions between different triad phases and probability distribution of triad phase precession frequencies.
\noindent ii) Joint probability distributions between precession frequencies and energy content and also between precession frequencies and growth rate of energy transfers. Second, in order to better understand the connections between these statistical properties and the presence of shock-like structure in the real space, we will study the effect of the introduction of a quenched disorder given by the projection of the dynamics on a fractal Fourier set, where preliminary results show a strong depletion of shocks \cite{michele2015} with reduced system fractal dimension. We will complement these studies with dynamical-system interpretations in terms of correlations and synchronisation within the state space.

\section{Burgers equation in one dimension and less}
\label{sec:Fractal_burg}

\begin{figure*}[tbh]
\resizebox{0.53\textwidth}{!}{
  \includegraphics{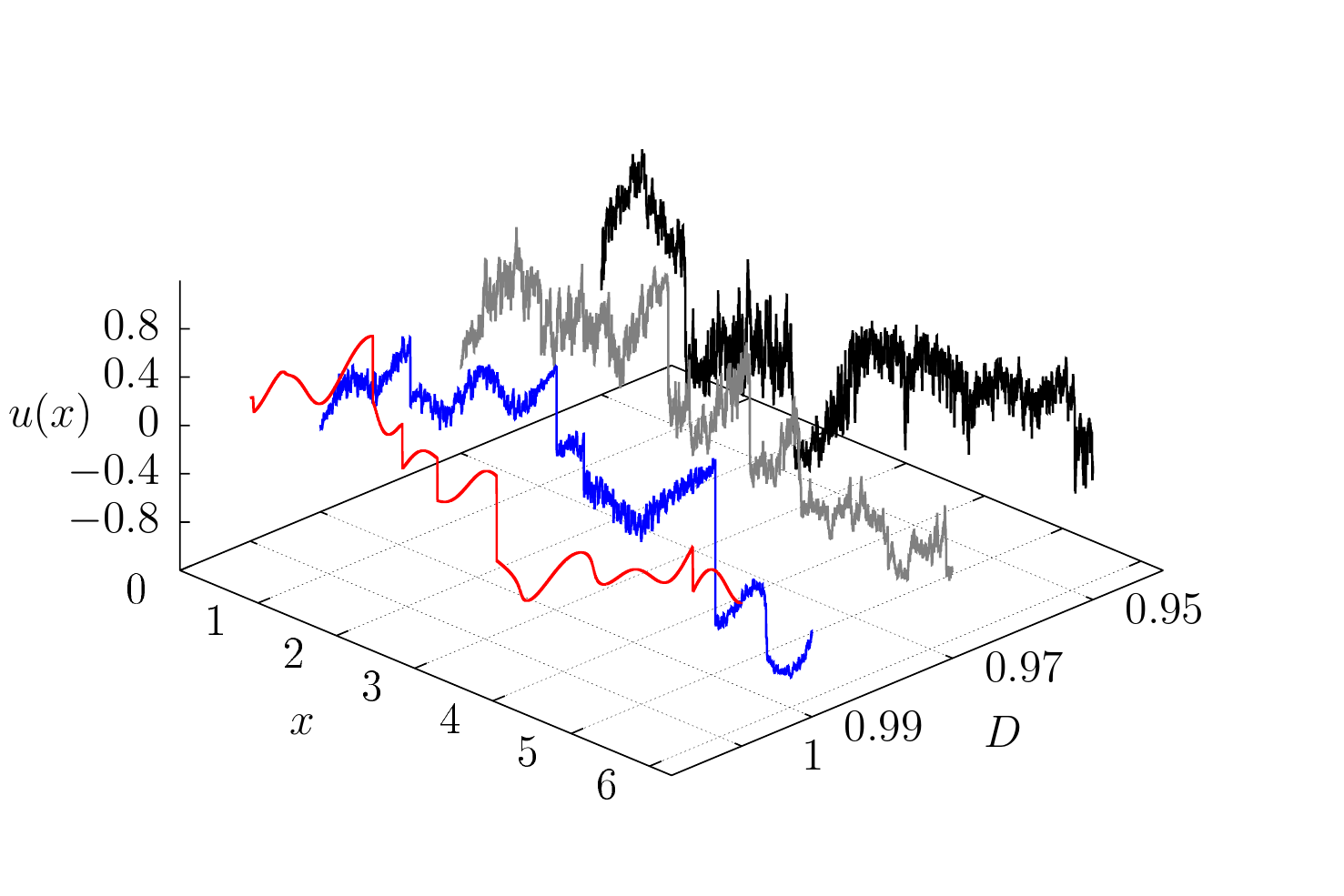}}
\resizebox{0.47\textwidth}{!}{
  \includegraphics{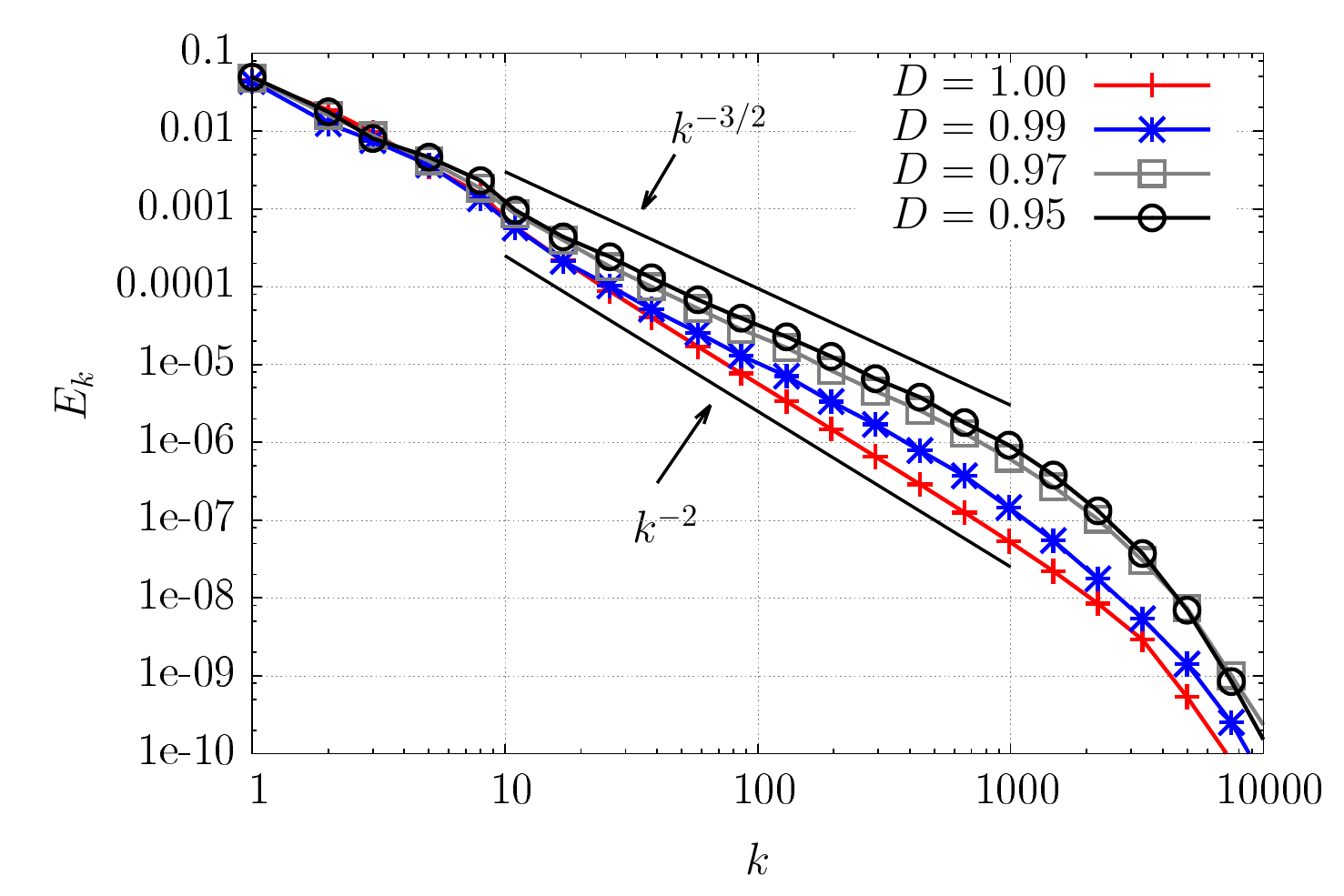}}
\caption{Left: Snapshots of real space velocity profiles as a function of space and fractal Fourier dimension: $D=1$ (red), $D=0.99$ (blue), $D=0.97$ (grey) and $D=0.95$ (black).  Right: Mean energy spectra in log-log scale, obtained for the same four fractal dimensions $D$ and using the same colour code as in the left panel. The black solid line is the scaling slope $k^{-2}$ produced by the undecimated ($D=1.0$) Burgers equation.}
\label{fig:burg_r_f}
\end{figure*}

Let us start from equation (\ref{eq:burgers}) written in Fourier space:
\begin{equation}
\frac{\partial \hat {u}_k}{\partial t} = -\frac{ik}{2} \sum_{k_1}\, \hat {u}_{k_1} \hat {u}_{k-k_1} - \nu k^2 \hat {u}_k + \hat {f}_k,
\label{eq:F_Burgers}
\end{equation}
where energy is dissipated from the viscous term and injected by the external forcing. The core of the dynamics is in the quadratic  non-linear convolution, a term that globally conserves energy by redistributing it amongst Fourier modes. Let us define the stationary Energy spectrum as:
\begin{equation}
\left <\hat {u}_{k}\hat {u}_{k'} \right > = E_{k} \delta(k+k'),
\label{eq:def_spec} 
\end{equation}
where $\left < \bullet \right >$ denotes an average over the statistically-steady ensemble. The energy flux  in the positive direction across a wavenumber $k$ is defined as $\Pi(k) \equiv -\frac{d}{dt}\left(\sum_{k_1=0}^k E_{k_1}\right)$. From (\ref{eq:F_Burgers}) and (\ref{eq:def_spec}) it is easy to realise that the contribution from the nonlinear interactions to this energy flux is given by the ``triadic terms'':
\begin{equation}
\label{eq:flux}
\Pi(k) = \sum_{k_1=\,1}^k \sum_{k_2=-\infty}^\infty \,  k_1 \, \Im \left \{\left < \hat {u}_{k_1} \hat {u}_{k_2}\hat {u}_{k_1 + k_2}^*\right >  \right \},
\end{equation}
where $\Im \{z\}$ denotes the imaginary part of $z$. It is important to point out that  the energy exchange produced by the nonlinear term inside a single triad is conservative. Indeed, the sum of the energy variation at three wave-numbers with  $k_1+k_2=k_3$  is always zero. This implies that the energy flux across a scale $k$ is only given by the energy exchanges inside triads which have at least one wave number higher than $k$ and one smaller than $k$. Hence, $\Pi(\infty)$ is always equal to zero \cite{Kraichnan1967,Kraichnan1971,Rose1978,Ohki_kida1992}. In particular, at those scales where neither the forcing mechanism nor the viscous term play any relevant role, the dynamics are fully dominated by the non-linear evolution, the so-called ``inertial-range'' of scales. For smaller and smaller viscosity, the solution of Burgers equations develops a velocity configuration with sharper and sharper shock-like structures (see fig.~\ref{fig:burg_r_f}), characterised by a power-law spectrum  $E(k) \propto k^{-2}$ (see \cite{becreview}).  

\subsection{Phases and fluxes}
\label{sec:maths}

Focusing on the non-linear term of the Burgers equation (\ref{eq:d-burgers}) and making the following substitution
\begin{equation}
\label{eq:ampl_phase_rep}
\hat u_k(t) = a_k(t) e^{i \phi_k(t)}, \qquad a_k(t) \equiv |\hat u_k(t)|\,, 
\end{equation}
and recalling the definition of triad phase
\begin{equation}
\label{eq:triad_phase}
\varphi_{k_1,k_2}^{k_3}(t) = \phi_{k_1}(t) + \phi_{k_2}(t) - \phi_{k_3}(t),
\end{equation}
we get the evolution for the amplitude $a_k$ and phase $\phi_k$ of the surviving modes:
\begin{eqnarray}
\hspace{-0.1cm}
\label{eq:ampl_Burg}
\dot{a}_k &=& \frac{k}{2} \, \theta_k \sum_{k_1,k_2} a_{k_1} a_{k_2} \,\theta_{k_1} \theta_{k_2} \sin(\varphi_{k_1,k_2}^{k}) \, \delta_{k_1+k_2,k} \, , \\
\label{eq:phas_ampl_Burg}
\dot{\phi}_k &=& -\frac{k}{2} \, \theta_k \sum_{k_1,k_2} \frac{a_{k_1} a_{k_2}}{a_k} \, \theta_{k_1} \theta_{k_2}\cos(\varphi_{k_1,k_2}^{k}) \, \delta_{k_1+k_2,k}.
\end{eqnarray}

Note that the factor $\theta_k$ in the latter equations is the projector in front of the nonlinear term in eq.~(\ref{eq:d-burgers}), while $\theta_{k_1}$ and $\theta_{k_2}$ represent the fact that also the other two modes satisfying the triadic condition are decimated. 

It is evident from eqs. (\ref{eq:triad_phase}), (\ref{eq:ampl_Burg}) and (\ref{eq:phas_ampl_Burg}) that $\varphi_{k_1,k_2}^{k_3}$, along with the $a_k$, satisfy a closed system of equations which determines the nonlinear dynamics. In particular, the individual phases $\phi_k$ can be obtained by quadratures from eq. (\ref{eq:phas_ampl_Burg}).

Applying substitutions (\ref{eq:ampl_phase_rep})--(\ref{eq:triad_phase}) to the velocity field in the energy flux equation, eq.~(\ref{eq:flux}), we get:

\begin{align} 
\nonumber
\Pi(k) = &\sum_{k_1=1}^{k} \sum_{k_2=-\infty}^\infty \left\langle {a}_{k_1} {a}_{k_2} {a}_{k_1+k_2}  \sin(\varphi_{k_1,k_2}^{k_1+k_2}) \right\rangle \\ 
\label{eq:flux_sin}
& \times k_1 \theta_{k_1} \theta_{k_2} \theta_{k_1 + k_2}.
\end{align}
As expected, the energy flux depends on the phases via $\varphi_{k_1,k_2}^{k_1+k_2}$ only. The crucial observation is that the sign of the contributions to this energy flux depends only on the values of the triad phases, because all other factors appearing in (\ref{eq:flux_sin}) are positive, including the real amplitudes $a_k$. Since the sign of the energy flux determines the cascade direction, it is very important to analyse these contributions in detail.

Define $k_3$ by the triad relation $k_1 + k_2 = k_3$. Following an  analogous method to the detailed energy balance of triads \cite{Kraichnan1967,Kraichnan1971,Rose1978,Ohki_kida1992}, we deduce that non-zero contributions to the energy flux (\ref{eq:flux_sin}) come from triads that have at least $1$ mode within the interval $[-k,k]$ and at least $1$ mode outside it.  
Combining these contributions we obtain
$$\Pi(k) = \sum_{k_1=1}^k \sum_{k_3=k+1}^{\infty} 2 k_1 \,\left\langle {a}_{k_1} {a}_{k_2}  {a}_{k_3}  \sin(\varphi_{k_1,k_2}^{k_3})\right\rangle\,,$$
where we have used the reality of the original field, which implies $\phi_{- k_j} = - \phi_{k_j}$ and $a_{- k_j}=a_{k_j}$. We remark the wavenumber ordering: $0 < k_1 < k_3, \quad 0 <  k_2 < k_3$ and $k_3 = k_1 + k_2$.

In summary, the sign of each individual triad's contribution depends solely on the sign of $\sin(\varphi_{k_1,k_2}^{k_3})$, where $k_1,k_2,k_3$ is the ``ordered'' version of that triad, with $k_3=k_1+k_2$ and $k_j > 0$. As a result, when $\varphi_{k_1, k_2}^{k_3}$ is close to $\frac{\pi}{2}+ 2n\pi$, $n \in \mathbb{Z}$, the direct-cascade energy flux is maximised (at fixed amplitudes). In Section \ref{sec:phase} we present statistical analyses that show that triad phases tend to cluster near these values, involving in addition a synchronisation amongst different triads, so that several terms contribute with a positive sign, leading to direct-cascade flux whose intensity depends on the dimension of the Fourier fractal set where modes live. 

\subsection{Burgers equation on fractal Fourier sets}

In order to study the effects of the quenched noise on the triadic interactions in Fourier space we use a projection method to fractal sets.
This projection method was originally introduced in \cite{frisch2012} to study the inverse energy cascade in 2D Navier-Stokes equations
 (see \cite{rayreview} for a review). Later, it was exploited to study the statistical properties 
of small-scale fluctuations 
in the 3D Navier-Stokes equations \cite{luca2015} and 1D Burgers equation \cite{michele2015}. In these two papers, a tendency towards more regular Gaussian statistics for the small-scale velocity field was observed at lower values of fractal dimension $D$, indicating that the disorder leads to an energy transfer regime characterised by self-similar fluctuations.

The projection on a fractal Fourier set is done using a decimation operator $P_D$ acting on the field $u(x,t)$ as
\begin{equation}
 \label{eq:projector}
v(x,t) = P_D u(x,t) =\sum_{k} e^{i k x}\theta_{k} \hat {u}_{k}(t).
\end{equation}
Here $\theta_{k}$, $k \in \mathbb{N}$ are independently chosen random numbers such that $\theta_{k} = 1$, with probability $h_k$ and $\theta_k = 0$, with probability $1-h_k$. We also impose the condition that $\theta_{k} = \theta_{-k}$ to ensure the reality of the field at all times. Choosing $h_k = k^{D-1}$, with $0<D\le 1$, 
we introduce a quenched disorder which randomly suppresses
modes on the Fourier lattice and ensures that on average we have 
$N(k)\propto k^D$ surviving modes inside an interval of length $k$ around the origin.
Applying this projector to the Burgers equation, we can then write the decimated PDE as
\begin{equation}
\frac{\partial v}{\partial t} + \frac{1}{2}P_D\frac{\partial v^2}{\partial x} = \nu\frac{\partial^2 v}{\partial x^2} + P_D f.
\label{eq:d-burgers}
\end{equation}
{We performed a set of numerical simulations of eq.~(\ref{eq:d-burgers}) by changing the dimension between $D=1$ and $D=0.95$. We chose the forcing to be Gaussian and white-in-time
\begin{table}
\caption{ $D$: dimension of the Fourier Fractal set where modes live. $N$: number of collocation points. $\%(D)$: percentage of decimated wave numbers, where the first value is related to the lower resolution used while the second value is related to the higher resolution. $\nu$: value of the 
kinematic viscosity. $k_f$: forced wavenumbers. $\delta t$: Numerical Time step used in the temporal evolution.}
\label{tab:1}
\begin{tabular}{cccccc}
\hline\noalign{\smallskip}
$D$ & $N$ & $\%(D)$ & $\nu$ & $k_f$ & $\delta t$\\
\noalign{\smallskip}\hline\noalign{\smallskip}
1 & $2^{16} - 2^{19}$ & $0$ & $8 \cdot 10^{-5}$ & $[1: 8]$ & $5.5 \cdot 10^{-5}$\\
0.99 & $2^{16} - 2^{19}$  & $8 - 10$ & $2.5 \cdot 10^{-5}$ & $[1: 8]$ & $2.3 \cdot 10^{-5}$\\
0.97 & $2^{16} - 2^{19}$  & $23 - 27$ & $9 \cdot 10^{-6}$ & $[1: 8]$ & $2.0 \cdot 10^{-5}$\\
0.95 & $2^{16} - 2^{19}$  & $36 - 40$ & $5 \cdot 10^{-6}$ & $[1: 8]$ & $1.7 \cdot 10^{-5}$\\
\noalign{\smallskip}\hline
\end{tabular}
\end{table}
\begin{equation}
\langle {\hat f}(k_1,t_1){\hat f}(k_2,t_2) \rangle = 2f_0|k_f|^{-1}\delta(t_1 - t_2)\delta(k_1 + k_2),
\end{equation}
acting only at large scales, the forced wave-numbers $k_f$ are all the non-decimated modes in the range $[1:8]$. For the numerical simulation we used Adams-Bashforth schemes of fourth order, with a number of collocation points ranged between $N = 2^{16}$ and $N = 2^{19}$, and a time step $\delta t \sim 10^{-5}$. See Table~\ref{tab:1} for more details about the numerical data.
Figure \ref{fig:burg_r_f} shows a snapshot of the velocity field for four different values of fractal dimension $D$, together with the mean energy spectra. It is important to mention that in fig.~\ref{fig:burg_r_f} the spectra have no gaps because we have further performed an average over different quenched fractal masks \cite{michele2015}. As one can see, shocks are present for all fractal dimensions, while the smooth (for $D=1$) ramps connecting 
\begin{figure*}[tbh]
\resizebox{0.5\textwidth}{!}{
  \includegraphics{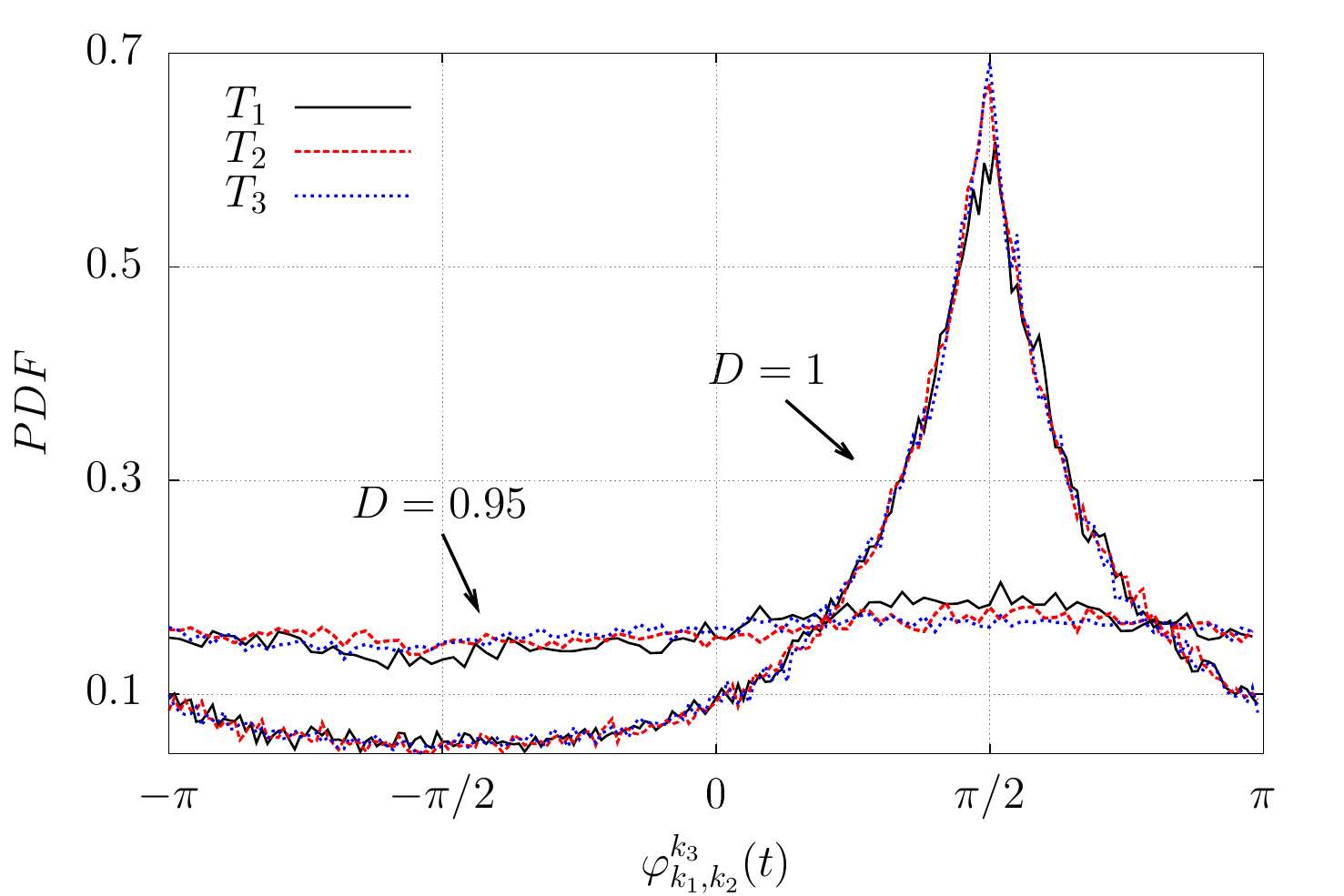}
}
\resizebox{0.44\textwidth}{!}{
  \includegraphics{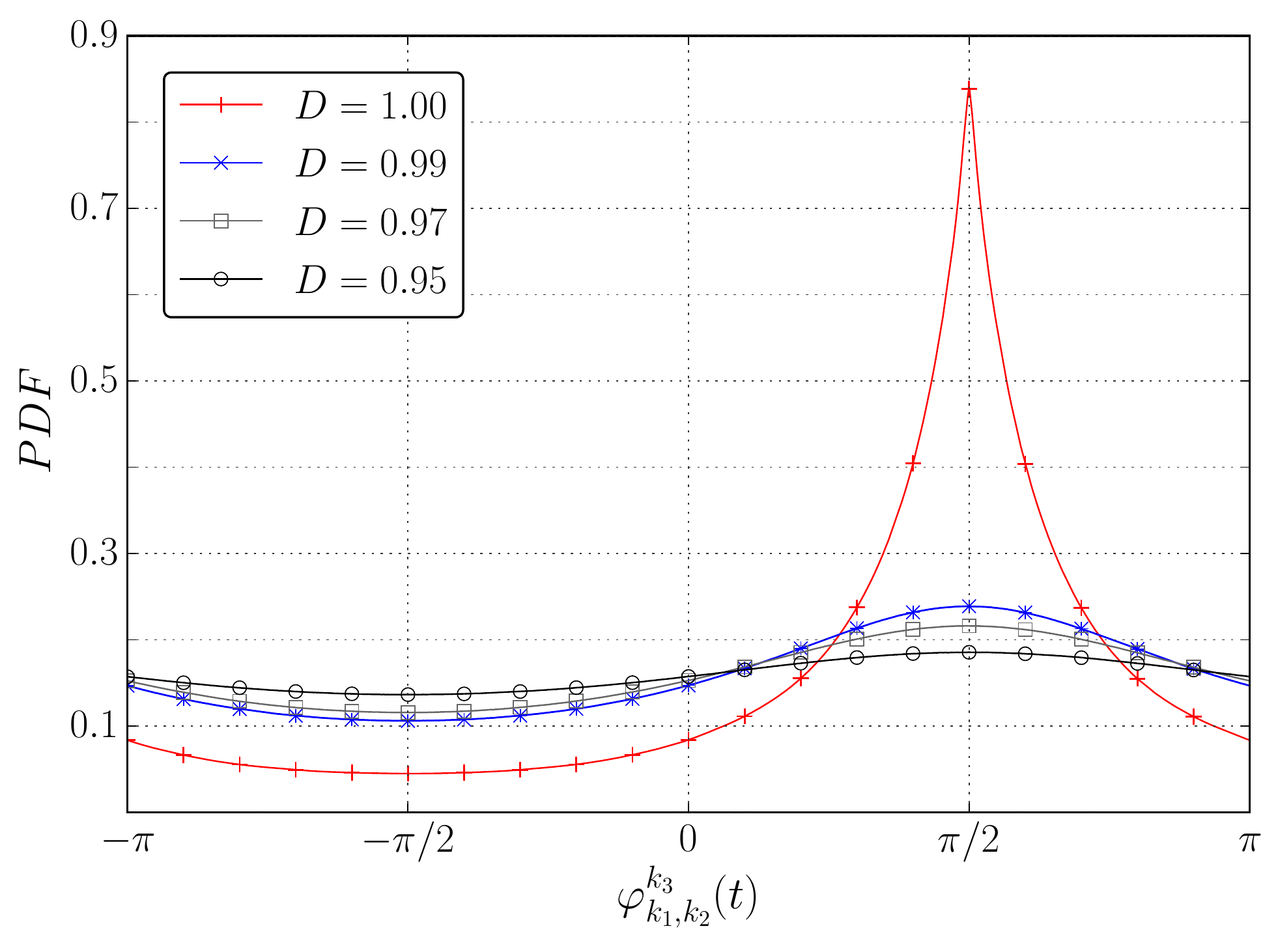}
}
\caption{Left: Histograms of triad phase (mod $2 \pi$ with offset $-\pi$) computed during the temporal evolution of different triads in the inertial range. $T_1: [k_1;k_2;k_3]=[100;150;250]$ (black solid line), $T_2: [k_1;k_2;k_3]=[200;250;450]$ (red dashed line), and $T_3: [k_1;k_2;k_3]=[300;350;650]$ (blue dotted line). To aid comparison, the undecimated case ($D = 1$) and decimated case ($D = 0.95$) are plotted together. Right: Same type of histograms but over all $160 801$ triads composed by wave numbers inside the range $100 \leq k \leq 1000$,  and for the same set of values of fractal Fourier dimension $D$ and colour code as in fig.~\ref{fig:burg_r_f}.}
\label{fig:phase_PDF}
\end{figure*}
two shocks develop small-scale fluctuations that become more and more pronounced as the fractal dimension $D$ decreases. It is important to remark that even if the disorder produces a non-trivial change in the spectrum, a power-law behaviour is always present \cite{michele2015}. Similar results have also been observed in the three dimensional Navier-Stokes equations \cite{luca2015}. A scaling of $k^{-2}$ is recovered in the undecimated $D=1$ case.}\\

\section{Numerical results}

\subsection{Phase analysis}
\label{sec:phase}
\begin{figure*}[tbh]
\hspace{-0.4cm}
\resizebox{1.04\textwidth}{!}{
  \includegraphics{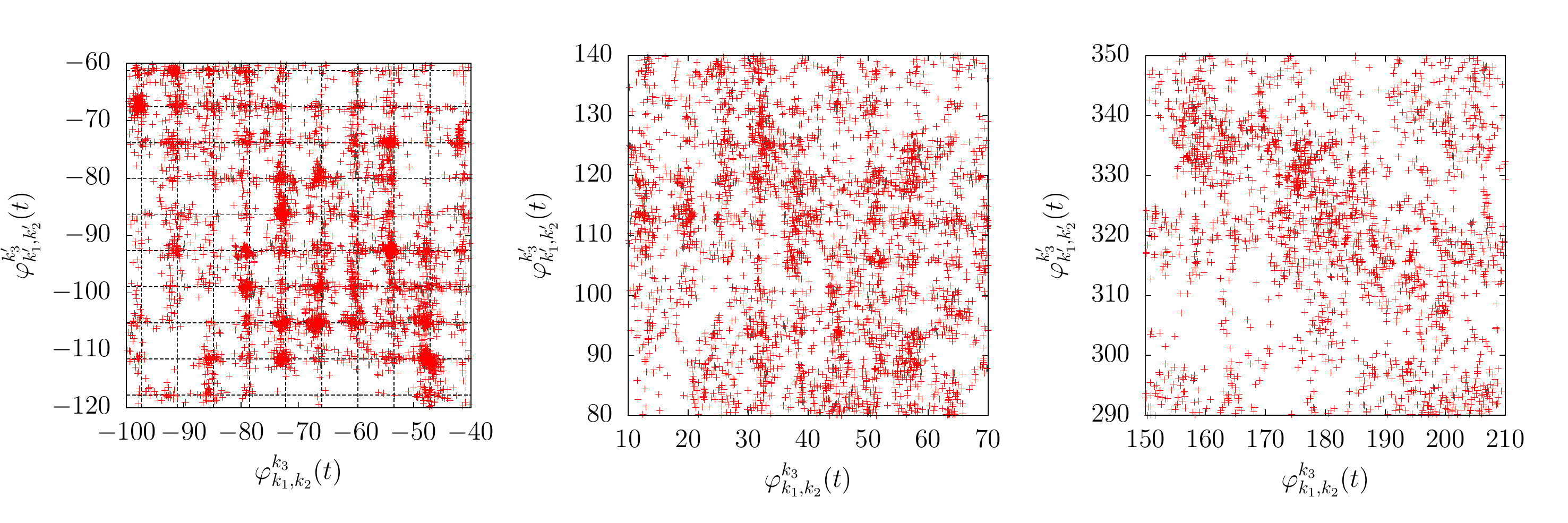}
}
\caption{Parametric plots of the phase evolution for two different triads in the inertial range: $T_1: [k_1;k_2;k_3]=[100;150;250]$ and $T_2: [k_1;k_2;k_3]=[200;250;450]$, showing synchronisation events near $\frac{\pi}{2}+2 n \pi$ , $n\in \mathbb{Z}$. Left: Fractal Fourier dimension $D = 1$. Grid lines denote triad phase values at $\frac{\pi}{2} + 2 n \pi$. Centre: Fractal Fourier dimension $D = 0.99$. Right: Fractal Fourier dimension $D = 0.95$.}
\label{fig:phase_scatter}
\end{figure*}

We present a phenomenological study of Burgers triad phase dynamics.
Let us first consider the case when we define all phases in the same
 periodic range, $\varphi_{k_1,k_2}^{k_3}(t) \in [-\pi, \pi]$, \emph{i.e.}, we omit the phase precessions for the moment. 
In fig.~\ref{fig:phase_PDF} (left panel) we show the PDF of $\varphi_{k_1,k_2}^{k_3}(t)$ for three different triads composed of wave numbers belonging to the inertial range and ordered in the usual way $0 < k_1 < k_2 < k_3$ with $k_3 = k_1+k_2$: 
\begin{align}
 \nonumber
 &T_1\rightarrow \,\,[k_1;k_2;k_3]=[100;150;250], \\ \nonumber
 &T_2\rightarrow \,\,[k_1;k_2;k_3]=[200;250;450], \\ \nonumber
 &T_3\rightarrow \,\,[k_1;k_2;k_3]=[300;350;650],
\end{align}
at $D=1$ and $D=0.95$. The undecimated case $D=1$ shows a more marked preference for triad phase alignments about $\pi/2$, which maximises the energy flux because flux contributions are proportional to $\sin (\varphi_{k_1,k_2}^{k_3})$ (see eq.~(\ref{eq:flux_sin})). The PDFs are peaked at $\pi/2$ and have a minimum at $3\pi/2$, which suggests that the energy flux is, on average, directed from large to small scales. This would be true provided one assumes that real amplitudes $a_k$ take higher or similar values when phases are near $\pi/2$ than when phases are near $3\pi/2$. This assumption is confirmed by simulations, so we have a direct cascade of energy as in 3D Navier-Stokes turbulence. On the other hand, for the case $D=0.95$ we no longer have a marked peak at $\pi/2$, especially when we consider the triad $T_3$ which is the one containing higher wavenumbers. This is the first evidence that the introduction of the quenched disorder strongly perturbs the coherence of the energy transfer mechanism, leading to a much slower direct cascade. In fig.~\ref{fig:phase_PDF} (right panel) we show a similar study but this time involving over $160$ thousand triads with wavenumbers in the range $[100, 1000]$. The features just discussed for the selected triads are again there, which shows that most triads are behaving in basically the same way.\\
A qualitative view of the correlation amongst different triad phases throughout the system's evolution is proposed in fig.~\ref{fig:phase_scatter}. Here a scatter plot between the phase of triads $T_1$ and $T_2$ is shown, where the phase angle is not constrained to be on a periodic domain. From the left panel (undecimated Burgers equations $D=1$) it is clear that a correlation is present between the two phases. A lattice-like distribution with the presence of many clusters indicates that for long periods the two phases are oscillating around a fixed value before jumping simultaneously toward other values. As expected, these clusters exist around values of $\pi/2 +  2 n \pi$, $n \in \mathbb{Z}$. This synchronization is directly linked to the results shown in fig.~\ref{fig:phase_PDF} where we saw a peak in the PDF for phase values at $\pi/2$. Since most triad phases spend most of their time near $\pi/2$, it is natural that several triads will be simultaneously near those values for some of the time. In contrast, results for $D=0.99$ and $0.95$ in fig.~\ref{fig:phase_scatter}, centre and right panels, show that the synchronization amongst different triad phases is washed out when the quenched disorder is introduced. 

Remarkably, synchronisation of triad phases has a direct effect on the energy flux. The more triads are synchronised near $\pi/2 +  2 n \pi$, the more positive terms add together to contribute to the flux in eq.~(\ref{eq:flux_sin}), leading to a large flux. Thus, loss of synchronisation diminishes the flux via cancellations of positive and negative terms.

\subsection{Precession analysis}

The ``drift'' or ``precession'' of triad phases made evident from fig.~\ref{fig:phase_scatter} suggests that we study the distribution of precession frequencies of the various triad phases in the system, as this distribution can have an impact on the energy fluxes via precession resonance \cite{bustamante14}.
In order to analyse the phase-precession statistical and dynamical properties it is important to filter out 
the high-frequency fluctuations introduced by the evolution of each of the phase variables 
contributing to the triad phase. Therefore, let us  
define the triad phase precession averaged on a time window $\Delta t$ as
\begin{equation}
\Omega^{k_3}_{k_1 k_2}(t,\Delta t) \equiv \frac{1}{N}  \sum_{t_i=t}^{t+\Delta t} \dot{\varphi}_{k_1 k_2}^{k_3}(t_i)\,,
\label{eq:num_press}
\end{equation}
where $t_i$ denote data points from the numerical integration and the time window, $\Delta t = N \delta t$, where $\delta t$ is the numerical time step and $N$ is a large enough integer to remove the noise at the dissipative scales. But $\Delta t$ should be chosen small enough not to kill all temporal correlations in the system. In practice we choose $\Delta t$ of the order of the Kolmogorov time microscale $t_\eta \equiv \sqrt{\frac{\nu}{\varepsilon}} \approx 10^{-2}$. 
The instantaneous phase precession, $\dot{\varphi}^{k_3}_{k_1 k_2}(t)$, is defined in terms of the finite-difference increments of individual phase derivatives $\dot{\phi}_k(t)$:
\begin{equation}
\dot{\phi}_k(t_i) = \frac{\phi_k(t_i + \delta t) - \phi_k(t_i)}{\delta t} = \frac{1}{\delta t} \arg \left [ \frac{\hat{u}_k(t_i+ \delta t)}{\hat{u}_k(t_i)} \right ] .
\label{eq:phase_diff}
\end{equation}

\begin{figure*}
\begin{center}
\resizebox{0.93\textwidth}{!}{
  \includegraphics{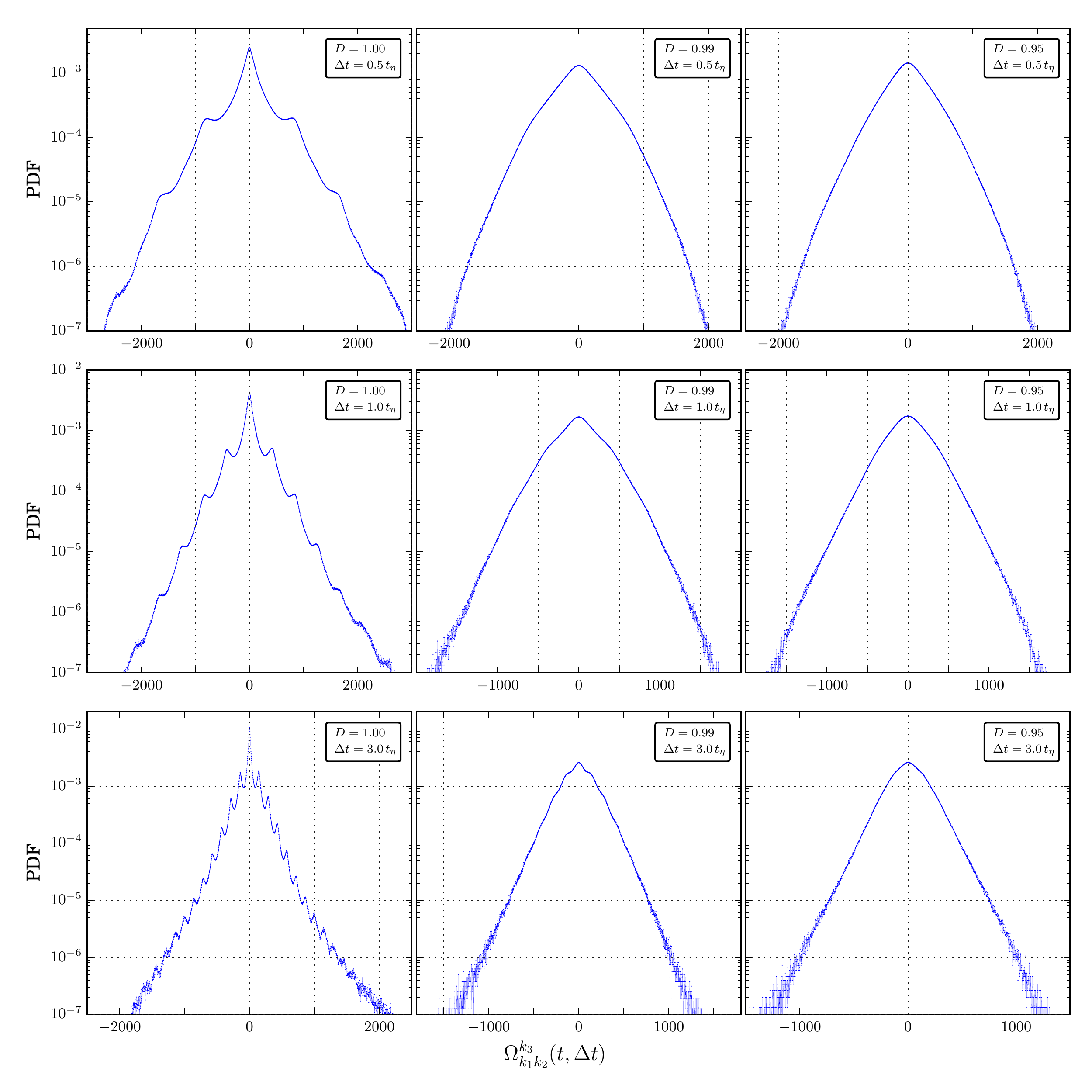}}
\end{center}
\caption{Probability density functions (PDFs) of the time-averaged triad precessions $ \Omega^{k_3}_{k_1 k_2}(t,\Delta t)$ (over time window $\Delta t$, see eq.~(\ref{eq:num_press})). The PDFs are computed over all $160,801$ triads composed by wave numbers inside the range $100 \leq k \leq 1000$ and for the temporal evolution $0 \leq t \leq 10^3 \,t_\eta$. Top row: $\Delta t  = 1000 \, \delta t = 0.5 \,t_\eta$. Middle row: $\Delta t  = 2000 \, \delta t = 1.0 t_\eta$. Bottom row: $\Delta t  = 6000 \, \delta t  = 3.0 \,t_\eta$. The three panels in each row represent different fractal dimensions. Left: $D = 1$. Centre: $D = 0.99$. Right: $D = 0.95$. }
\label{fig:pdf_prec}
\end{figure*}

The probability density functions (PDFs) of $\Omega^{k_3}_{k_1 k_2}(t, \Delta t)$ are produced by considering all $160 801$ triads constructed from modes inside an interval in the inertial range: $k \in I_k = [100, 1000]$. In fig.~\ref{fig:pdf_prec} we show the corresponding PDFs for three different time windows: $\Delta t  = 1000 \, \delta t  = 0.5 \,t_\eta$ (top row); $\Delta t  = 2000 \, \delta t  =1.0 \,t_\eta$ (middle row); $\Delta t  = 6000 \, \delta t = 3.0 \,t_\eta$ (bottom row). Furthermore, we also compare the results obtained from the original undecimated ($D=1$) Burgers equation (left), with two different decimated systems: $D=0.99$ (centre) and $D=0.95$ (right). 
We first note the appearance of secondary peaks in the PDFs, corresponding to non-zero precession frequency values. 
The number of peaks increases as the time window $\Delta t$ increases because the triad phase has had enough time to make several jumps between adjacent trapping regions, namely from $\pi/2 + 2n\pi$ to $\pi/2 + 2n'\pi$, with $|n-n'|=1$, in apparently random sequences (figure not shown, but the jumpy behaviour is evident from the scatter plots of fig. \ref{fig:phase_scatter}). 
Similarly, the peaks become more pronounced as $\Delta t$ increases because the longer averaging window leads to a suppression of oscillations in $\varphi_{k_1 k_2}^{k_3}$ around the trapping regions $\pi/2 + 2n\pi$. This explanation also applies to the secondary peaks where the time window $\Delta t$ includes both the oscillations of $\varphi_{k_1 k_2}^{k_3}$ referred to above but also a number of jumps in $\varphi_{k_1 k_2}^{k_3}$ towards other trapping regions. As $\Delta t$ is increased the contributions of the oscillations are suppressed but the contributions from the jumps will accumulate. The second interesting feature is the sudden disappearance of the secondary precession-frequency peaks as we decimate the system. At $D=0.99$ they are barely detectable and at $D=0.95$ they have been wiped out. 
This is in line with the results presented in figs.~\ref{fig:phase_PDF} and~\ref{fig:phase_scatter}, 
showing a transition from a structured phase evolution for $D=1$ to a disordered distribution when decimation is applied. \\
In order to explore the energy flux it is also important to examine the correlation between evolution of triad precession and triad amplitude. For the same subset of triads, $k_1 + k_2 = k_3$ in the range $I_k$, we computed two different joint PDFs, corresponding to the correlation amongst the triad precession $\Omega^{k_3}_{k_1 k_2}(t, \Delta t)$ and the temporal average, over $\Delta t$, of: i) the modulus of the product of the mode amplitudes,
\begin{equation}
\left< \left| a_{k_1} a_{k_2} a_{k_3}\right|\right>_{\Delta t},
\label{eq:amp_triad}
\end{equation}
and ii) the modulus of the product of the logarithmic time derivatives of the mode amplitudes,
\begin{equation}
\left<\left| \frac{\dot{a}_{k_1} \dot{a}_{k_2} \dot{a}_{k_3}}{ a_{k_1} a_{k_2} a_{k_3} } \right| \right>_{\Delta t}.
\label{eq:ampdot_triad}
\end{equation}
The motivation for choosing these quantities is that they provide a fair comparison weight, amongst different triads, with respect to energy content i) and flux of energy ii).

\begin{figure*}
\begin{center}
\resizebox{1.0\textwidth}{!}{
  \includegraphics{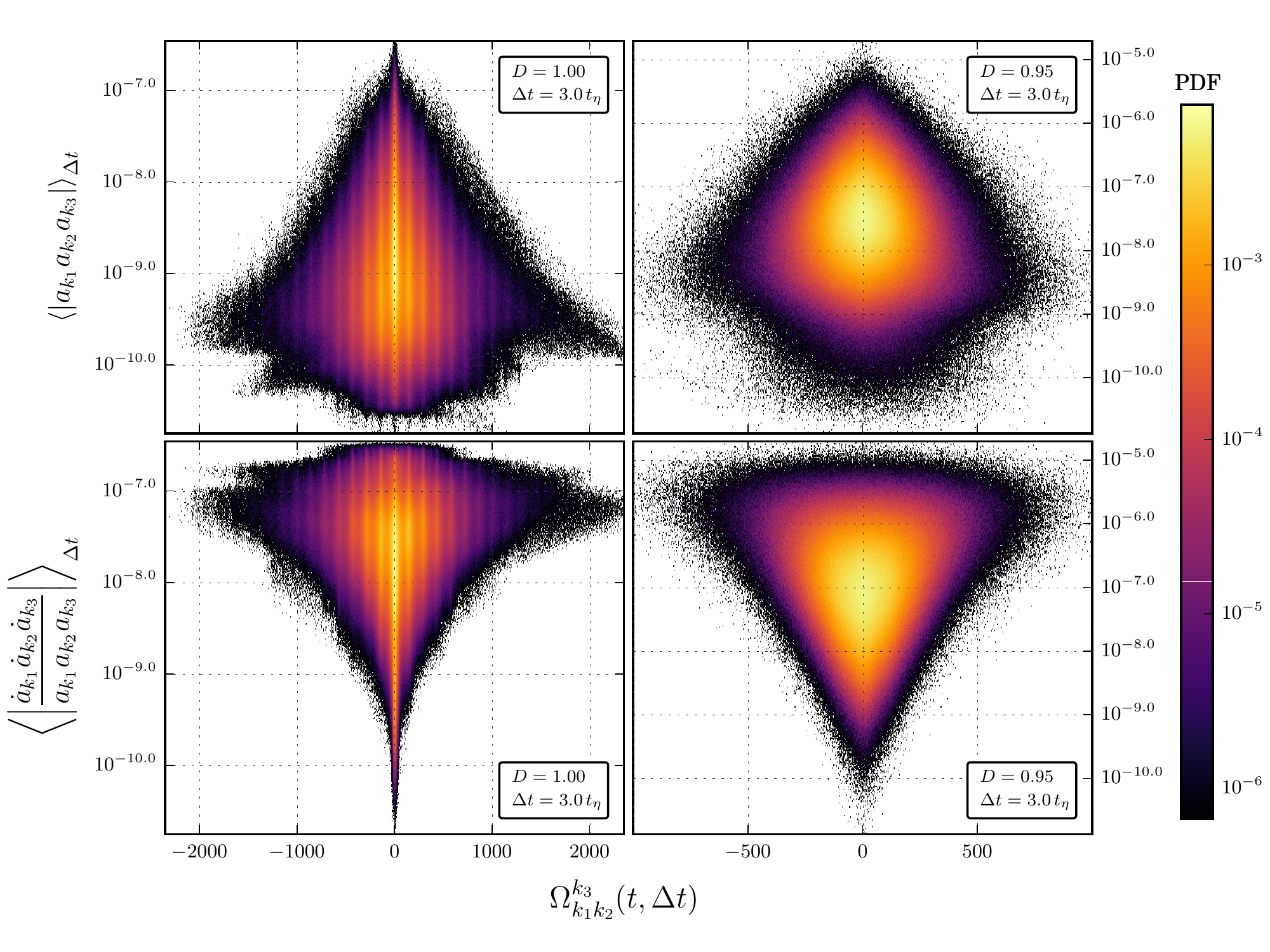}
}
\end{center}
\caption {Joint PDFs for time window $\Delta t  = 3.0 \,t_\eta$, for the same set of triads as in fig.~\ref{fig:pdf_prec}.  Top row: PDFs as functions of triad precession frequency and product of triad mode amplitudes. Bottom row: PDFs as functions of triad precession frequency and product of logarithmic time derivative of triad mode amplitudes. Left: $D = 1$. Right: $D = 0.95$.}
\label{fig:2D_histo}
\end{figure*}

In fig.~\ref{fig:2D_histo} we show the resulting joint PDFs for a time window $\Delta t  = 3.0 \,t_\eta$, together with its dependency on the fractal dimension $D$.
From these PDFs one can again detect the presence of secondary correlation peaks in the undecimated case only ($D=1$). The introduction of the quenched disorder destroys the cross-correlations amongst amplitudes 
and phase precessions.

A closer inspection of fig.~\ref{fig:2D_histo} reveals a connection with a dynamical systems approach, which deserves some mention here. By definition, a critical point of system (\ref{eq:phas_ampl_Burg}) is a point in the state space where \emph{all} time derivatives $\dot{a}_k$ and $\dot{\varphi}_{k_1 k_2}^{k_3}$ are equal to zero simultaneously, for all modes and triads. The most common type of critical point is the so-called ``unstable'': the state of the system may approach such a point along some directions, getting close to it but eventually separating from it along other directions. Thus, active energy exchanges are expected if the state of the system gets close to an unstable critical point.
 The evidence from fig.~\ref{fig:2D_histo} is that the state of the Burgers system visits unstable critical points regularly during the evolution: notice the tail near $\dot{a}_k =0, \,\dot{\varphi}_{k_1 k_2}^{k_3}=0$ at the lower part of the bottom left panel. Moreover, this tail is correlated, by definition, with the tail at the upper part of the top left panel, corresponding to events of maximum amplitude, \emph{i.e.}, high energy transfers. 
Finally, fig.~\ref{fig:phase_scatter} suggests that these critical points are distributed preferentially near values of the triad phases given by ${\varphi}_{k_1 k_2}^{k_3} = \pi/2 + 2 n \pi$. In contrast, in the fractal decimated case ($D=0.95$; see fig.~\ref{fig:2D_histo}, right panels) the above-mentioned tails lose coherence, which could mean that the state visits critical points less often or that less critical points exist. These and other implications will be studied in detail in subsequent work.

\section{Conclusions}
\label{sec:conclusions}

In this work we have presented numerical results about the evolution of the phases of the triadic structures responsible for the energy transport in the Fourier representation of 1D Burgers equation. Figure \ref{fig:phase_PDF} shows that the PDFs of these phases are characterised by marked peaks at $\pi/2 + 2 n \pi$, consistent with the presence of a well-defined energy flux directed towards the small scales. In addition, strong correlations between the phase evolution of different triads are presented in fig.~\ref{fig:phase_scatter}. These correlations hint to the entangled interactions and synchronisation developed by the Burgers nonlinearity amongst all the Fourier modes in order to provide the well known energy cascade.  

The PDF of precession frequencies, fig.~\ref{fig:pdf_prec}, shows a structure of secondary peaks with significant probability, circumscribed by a non-Gaussian (fat-tailed) envelope. This non-trivial feature motivates the search for higher-order precession resonances, responsible for perturbations in the energy flux across scales. These resonances occur when the precession frequency ${\Omega}_{k_1 k_2}^{k_3}$ of a triad phase coincides with a typical nonlinear frequency of the triad's mode amplitudes \cite{bustamante14}. Such a search requires the introduction of a \emph{tuning parameter}. Work in this direction will be reported elsewhere. 

Another important aspect of this paper consists of the study of the effects produced by the introduction of fractal-Fourier decimation for the 1D Burgers equation. We found clear evidence that decimation drives the system to a loss of temporal and spatial correlations among the surviving triad amplitudes and phases, as figs.~\ref{fig:phase_scatter}, \ref{fig:pdf_prec} and \ref{fig:2D_histo} show. In real space, as the dimension $D$ is lowered, the velocity field tends to develop fluctuations in the ramps between shocks. These increasing fluctuations could be linked to the loss of correlation/synchronisation amongst the triad amplitudes and phases. In particular, the significant correlation between triad precession frequencies and amplitude growth in the undecimated case ($D=1$) is lost as the dimension $D$ is lowered (see fig.~\ref{fig:2D_histo}). 

Looking forward, the dynamical systems point of view is worthy of further detailed investigation. We need to understand how phases organise and synchronise during the dynamical evolution, what is the distribution of critical points in the state space, and how these features change when we reduce the fractal dimension $D$. In particular, the alternation of trapping events and jumping events experienced by the triad phases in fig.~\ref{fig:phase_scatter} suggests the possibility to develop a model for phase evolution based on a biased random-walk with waiting times \cite{Montroll}, which in turn could explain the fat tails seen in fig.~\ref{fig:pdf_prec}. Finally, we stress that very similar studies can be carried on for the much more complicated case of turbulence in the three dimensional Navier-Stokes equations. Work in this direction will be reported elsewhere. \\ \\
BPM and MDB acknowledge support from Science Foundation Ireland under research grant number 12/IP/1491, and computational resources and support provided by the DJEI/DES/SFI/HEA Irish Centre for High End Computing (ICHEC) under class C project ndmat025c.
MB and LB acknowledge funding from the European Research Council under the European Union's Seventh Framework Programme, ERC Grant Agreement No 339032 and the INFN HPC initiative \emph{Zefiro}. MB acknowledges the kind hospitality of the Complex and Adaptive Systems Laboratory, School of Mathematics and Statistics, University College Dublin. We thank the COST-Action MP1305 for support.


%
%

\end{document}